\documentclass{knawproc}
\usepackage{epsfig}

\begin{document}

\begin{opening}

\title{Starbursts and the High-Redshift Universe}

\author{Timothy M. Heckman}
\addresses{%
  Department of Physics and Astronomy, Johns Hopkins University, Baltimore, MD
21218 USA\\
}

\end{opening}

\begin{abstract}
Starbursts are episodes of intense star-formation that occur in the
central regions of galaxies, and dominate the integrated emission
from the galaxy. They serve as local analogs of
the processes that were important in the origin and early evolution
of galaxies and in the heating and chemical enrichment of the
inter-galactic medium. They may also play an important role in the
AGN phenomenon.
In this contribution I review starbursts
from this broad perspective, with a specific focus on the use of
UV spectroscopic diagnostics that can be `calibrated' at low-redshift
and then applied at high redshift. From the analysis of the UV properties
of local starbursts we have learned: 1) dust dramatically affects our view of
high-mass star-formation 2) more metal-rich starbursts are redder and more
heavily extincted in the UV, more luminous, have stronger vacuum-
UV absorption-lines, and occur in more massive and optically-brighter host
galaxies 3) the strong interstellar absorption-lines directly
reflect the hydrodynamical consequences of the starburst. 
These results suggest that the high-
redshift `Lyman Drop-Out' galaxies are typically highly reddened
and extincted by dust (average factor of 5 to 10 in the UV), may
have moderately high metallicities (0.1 to 1 times solar?), are
probably building galaxies with stellar surface-mass-densities
similar to present-day ellipticals, and may be suffering
substantial losses of metal-enriched gas that can `pollute' the
inter-galactic medium. I also discuss UV observations of the nuclei
of type 2 Seyfert galaxies. These show that compact (100-pc-scale)
heavily-reddened starbursts are the source of most of the `featureless
continuum' in UV-bright Seyfert 2 nuclei, and are an energetically
significant component in these objects.

\end{abstract}

\section{Introduction}

Starbursts are sites of intense star-formation that occur in the
`circum-nuclear' (kpc-scale) regions of
galaxies, and dominate the integrated emission from the `host'
galaxy (cf. Leitherer et al 1991). The implied star-formation rates
are so high that the existing gas supply may sustain the starburst
for only a small fraction of a Hubble time (in agreement with
detailed models of the observed properties of starbursts, which
imply typical burst ages of-order 10$^{8}$ years). Both optical
objective prism searches and the IRAS survey have shown that
starbursts are major components of the local universe (cf. Huchra 1977;
Gallego et al 1995; Soifer et al 1987). Indeed, integrated over the local
universe, the total rate of (high-mass) star-formation in
circumnuclear starbursts is comparable to the rate in the disks of
spiral galaxies (Heckman 1997). Thus, starbursts deserve to be
understood in their own right.
 
Starbursts are even more important when placed in the broader
context of contemporary stellar and extragalactic astrophysics.
The cosmological relevance of starbursts has been dramatically
underscored by one of the most spectacular discoveries in years:
the existence of a population of high-redshift (z $>$ 2) star-forming
field galaxies (cf. Steidel et al 1996; Lowenthal et al 1997). The
sheer number density of these galaxies implies that they almost
certainly represent precursors of typical present-day galaxies in
an early actively-star-forming phase. This discovery therefore
moves the study of the star-forming history of the universe into
the arena of direct observations (Madau et al 1996), and gives
added impetus to the quest to understand local starbursts.

Starbursts may also play a vital role in the AGN phenomenon. Indeed,
perhaps the most important unanswered question concerning the AGN
phenomenon is the fundamental nature of the energy source. There
are sound theoretical arguments in favor of accretion onto
supermassive black holes (e.g. Rees 1984), and the observational
evidence that such `beasts' exist is growing (Kormendy \& Richstone
1995; Miyoshi et al 1995; Tanaka et al 1995). On the other hand,
circumnuclear starbursts can have bolometric luminosities that
rival even powerful QSO's (cf. Sanders \& Mirabel 1996 and
references therein), and there have been recurring suggestions that
such starbursts may play an important role in the Seyfert galaxy
phenomenon (e.g. Weedman 1983; Perry \& Dyson 1985; Terlevich \&
Melnick 1985; Norman \& Scoville 1988; Cid Fernandez \& Terlevich
1995).
 
Observations in the vacuum-UV spectral regime are crucial for
understanding local starbursts, for relating them to galaxies
at high-redshift, and for probing the `starburst-AGN connection'.
Only in this spectral regime can we clearly
observe the direct spectroscopic signatures of the hot stars that
provide most of the bolometric luminosity of starbursts
(e.g. Sekiguchi \& Anderson 1987; Fanelli, O'Connell, \& Thuan 1988;
Leitherer, Robert,
\& Heckman 1995). Moreover,
the vacuum-UV contains a wealth of spectral features, including the
resonance transitions of most cosmically-abundant ionic species
(cf. Kinney et al 1993).
These give UV spectroscopy a unique capability for diagnosing the
(hot) stellar population and the physical and dynamical state of
gas in starbursts and AGN.
 
Since ground-based optical observations of galaxies at
high-redshifts sample the vacuum-UV portion of their rest-frame
spectrum, we can not understand how galaxies evolved without
documenting the vacuum-UV properties of galaxies in the present
epoch. In particular, a thorough understanding of how to exploit
the diagnostic power of the rest-frame UV spectral properties of
local starbursts will give astronomers powerful tools with which
to study star-formation and galaxy-evolution in the early universe.

In the present paper I will first briefly review the observed UV
properties of starbursts in the local universe. I will then describe
the implications these results have for our understanding of star-forming
galaxies at high-redshift. Lastly, I will describe results from
our on-going program of UV imaging and spectroscopy of type 2 Seyfert
nuclei and discuss their implications for the `starburst-AGN connection'.

\section{UV Spectroscopy of Local Starbursts}

\subsection{Observational Overview}

\subsubsection{The UV Continuum: Probe of Dust}
 
The effect of dust on the UV properties of starbursts is profound.
Previous papers have established that various independent
indicators of dust extinction in starbursts
correlate strongly with one another. Calzetti et al (1994;1996)
show that the spectral slope in the vacuum-UV continuum (as
parameterized by $\beta$, where F$_{\lambda} \propto\
\lambda^{\beta}$) correlates strongly with the nebular extinction
measured in the optical using the Balmer decrement. Meurer et al
(1997;1998) show that $\beta$ also correlates well with the ratio of
far-IR to vacuum-UV flux: the greater the fraction of the UV that
is absorbed by dust and re-radiated in the far-IR, the redder the
vacuum-UV continuum. The interpretation of these correlations with
$\beta$ in terms of the effects of dust are particularly plausible
because the {\it intrinsic} value for $\beta$ in a starburst is a
robust quantity. Figures 31 and 32 in Leitherer \& Heckman (1995)
show that $\beta$ should have a value between about -2.0 and -2.6
for the range of ages, initial mass functions, and 
metallicities appropriate for
starbursts.
 
\subsubsection{The UV Lines: Probes of Gas \& Stars}
 
The vacuum-UV spectra of starburst are characterized by strong
absorption features. These absorption features can have three
different origins: stellar winds, stellar photospheres, and
interstellar gas.
 
Detailed analyses of HST and HUT spectra of starbursts show that
the resonance lines due to species with low-ionization potentials
(OI, CII, SiII, FeII, AlII, etc.) are primarily interstellar in
origin. In contrast, the resonance lines due to high-ionization
species (OVI, NV, SiIV, CIV) can contain significant contributions from
both stellar winds and interstellar gas, with the relative
importance of each varying from starburst to starburst (Conti et
al 1996; Leitherer et al 1996; Heckman \& Leitherer 1997;
Gonzalez-Delgado et al 1998a,b; Robert et al 1998). While the
oft-used `fiducial' UV spectrum of the starburst NGC 1741 clearly
shows the characteristic broad, blueshifted stellar wind profiles
in these high-ionization lines,
NGC 1705 is a counter-example in which the
interstellar medium actually dominates the CIV and SiIV absorption
(York et al 1990; Heckman \& Leitherer 1997; Sahu \& Blades 1997).
A comparison of the two starburst spectra is shown in Figure 1.
The most unambiguous detection of stellar photospheric lines in
starbursts is
provided by excited transitions, which are usually rather weak (cf.
Heckman \& Leitherer 1997).


\begin{figure}
\centerline{\epsfig{figure=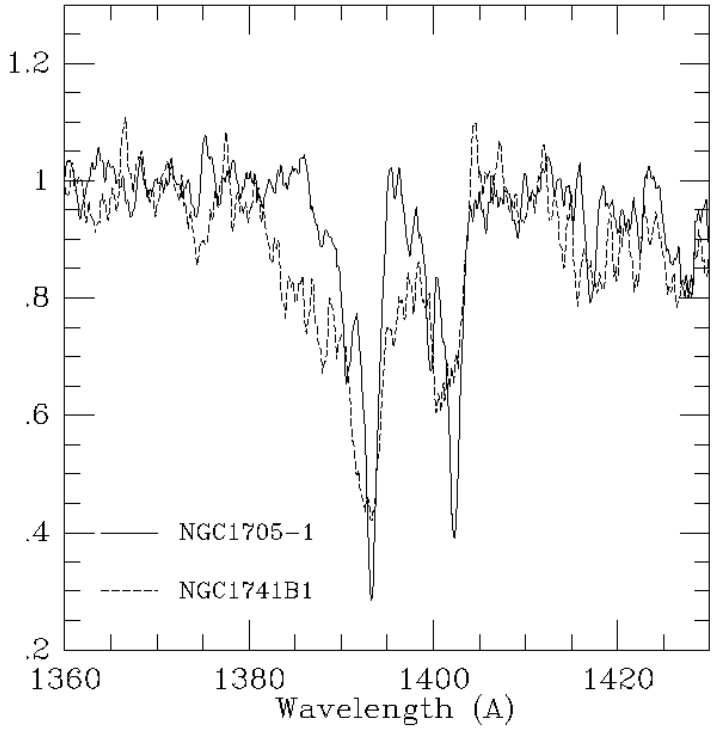}}
\caption{Spectral region around the SiIV$\lambda\lambda$1394,1403 doublet.
The starbursts NGC 1705
and NGC 1741 are compared. The spectra were normalized and the wavelength
 scales are in the restframe of each galaxy. Note the strong blue-shifted
wing on the SiIV profile in the NGC 1741 spectrum, indicating hot star
winds. The narrow doublet in NGC 1705 is instead largely due to
interstellar gas. Thus, even though the SiIV equivalent width is
roughly the same in both starbursts, spectra with adequate spectral
resolution and signal-to-noise are required to establish its physical
origin. The implications for the interpretation of spectra of galaxies
at high-redshift are obvious.}
\end{figure}

As a guide to the discussion to follow, in Figure 2, I present two
high signal-to-noise vacuum-UV spectra formed by averaging together
the IUE spectra of many starburst galaxies.
The first template
(`low-metallicity starburst') was formed from starbursts with
metallicity less than 0.4 solar (mean of 0.16 solar). The
second (`high-metallicity starburst') was formed from starbursts 
with metallicity greater than 0.5 solar
(mean of 1.2 times solar).
It is interesting
to note that the vacuum-UV spectra of starbursts are dominated by
strong {\it absorption}-lines, but the nebular {\it emission}-lines are very
weak (the reverse of the situation in the visible - cf. Leitherer,
1997).

\begin{figure}
\centerline{\epsfig{figure=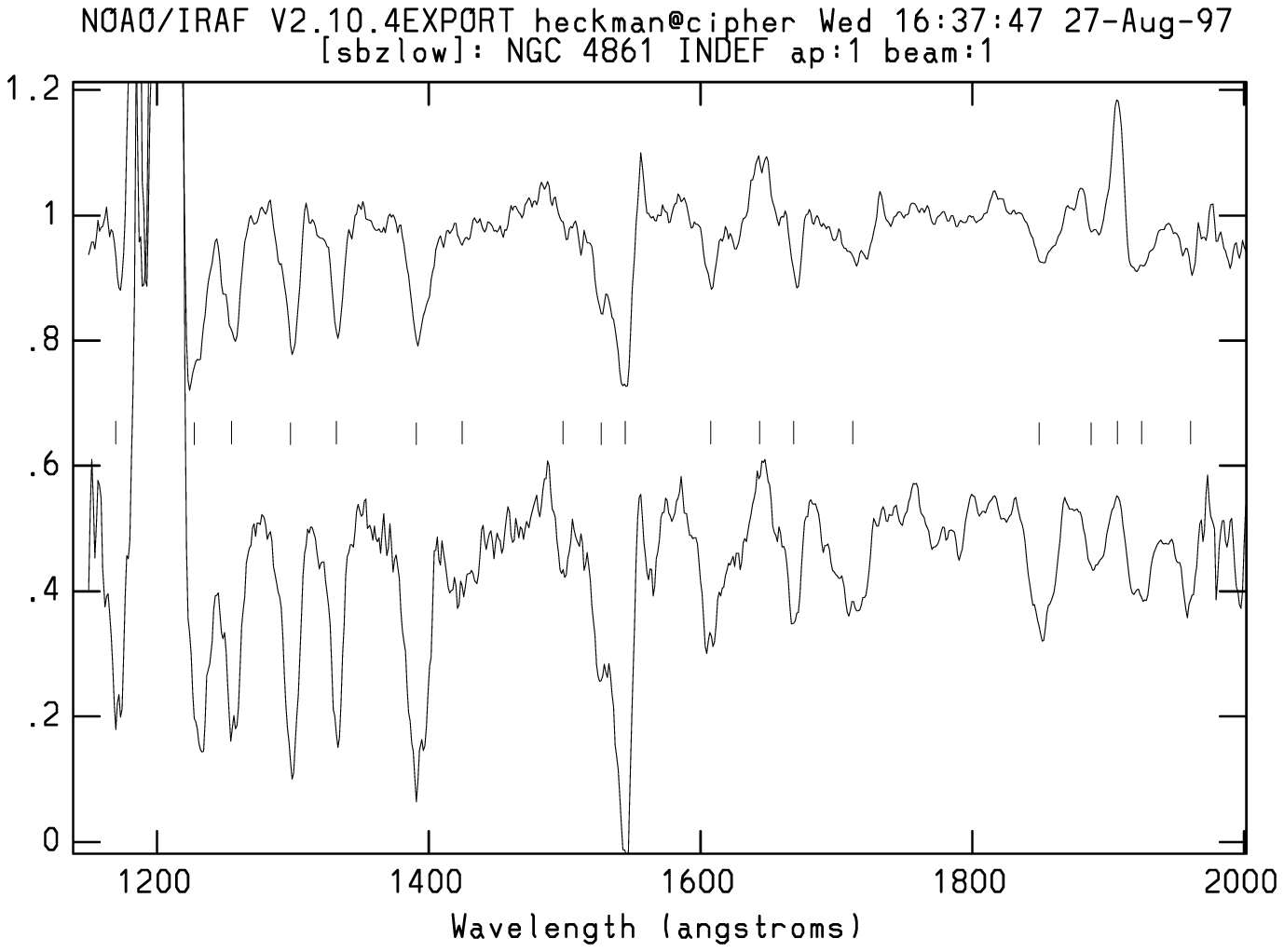}}
\caption{IUE spectra of local starbursts with low-metallicity
(top) and high-metallicity (bottom). Each spectrum is a weighted
average of the spectra of about 20 starbursts. The mean metallicities
are 0.16 solar (top) and 1.2 solar (bottom). A number of features
are indicated by tick marks and have the following identifications
(from left-to-right): CIII$\lambda$1175 (P), NV$\lambda$1240 (W),
SiII$\lambda$1260 (I), OI$\lambda$1302 plus SiII$\lambda$1304 (I),
CII$\lambda$1335 (I), SiIV$\lambda$1400 (W;I), SiIII$\lambda$1417 plus
CIII$\lambda$1427 (P), SV$\lambda$1502 (P), SiII$\lambda$1526 (I),
CIV$\lambda$1550 (W;I),
FeII$\lambda$1608(I), HeII$\lambda$1640 emission (W), AlII$\lambda$1671 (I),
NIV$\lambda$1720 (W), AlIII$\lambda$1859 (I;W), SiIII$\lambda$1892 (P),
CIII]$\lambda$1909 (nebular emission-line), FeIII$\lambda$1925 (P),
FeIII$\lambda$1960 (P). Here, I, P, and W denote
lines that are primarily of interstellar, stellar photospheric, or stellar
wind origin. The strong emission feature near 1200 \AA\ is geocoronal
Ly$\alpha$.}
\end{figure}

\subsection{Lessons Learned}

We (Heckman et al 1998 - hereafter H98) have just completed an
analysis of the vacuum-UV spectroscopic properties of a large
sample of starburst galaxies in the local universe using the data
archives of the International Ultraviolet Explorer (IUE) satellite.
Taken together with the results of previous 
spectroscopic studies of starbursts in the
vacuum-UV, the principal lessons we have learned are as follows:\\

{\bf First, UV spectra of adequate spectral resolution and signal-to-noise
are required to ascertain the origin of the observed absorption-lines.}

The data on NGC 1705 (Heckman \& Leitherer 1997) highlight the
difficulty in using the UV spectra of galaxies to deduce their
stellar content: data of relatively high spectral resolution and
signal-to-noise are needed to reliably isolate the stellar and
interstellar components. Simply measuring the equivalent widths of
the lines is not enough: it is the {\it profile shapes} that
contain the key information (see Figure 1).\\
 
{\bf Second, dust has a profound effect on the emergent UV
spectrum.}

This point has been discussed in section $2.1.1$ above, and further
details may be found in Meurer et al (1997;1998). Here, I just want
to mention some of the systematic ways in which the extinction
correlates with other fundamental properties of the starburst.

The amount of dust-extinction and reddening is well-
correlated with metallicity, as shown in Figure 3.
At low metallicity ($<$10\% solar) a significant fraction of the
intrinsic vacuum-UV actually escapes the starburst
(L$_{IR}$/L$_{UV}$ $\sim$ unity), and the vacuum-UV colors are
consistent with the intrinsic (unreddened) colors expected for a
starburst population ($\beta \sim$ -2.3). In contrast, at high
metallicities ($>$ solar) 90\% to 99\% of the energy emerges
in the far-IR (L$_{IR}$/L$_{UV}$ = 10 to 100) and the vacuum-UV
colors are very red ($\beta \sim$ 0). Storchi-Bergmann, Calzetti,
\& Kinney (1994) had previously noted the correlation between
metallicity and UV color. 

 
\begin{figure}
\centerline{\epsfig{figure=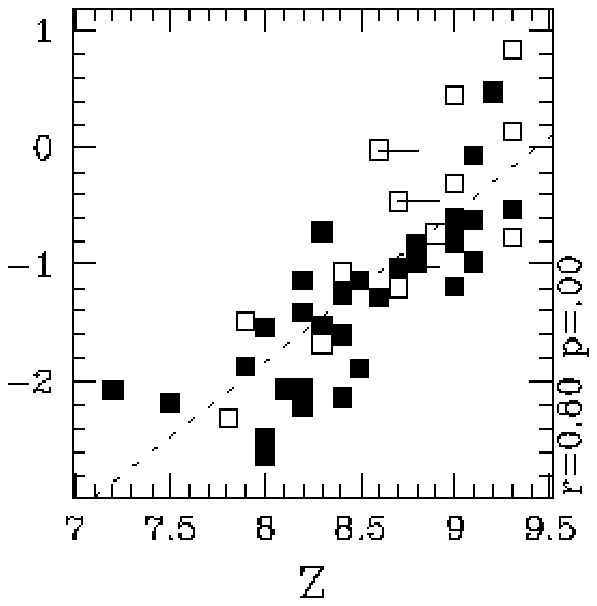}
\epsfig{figure=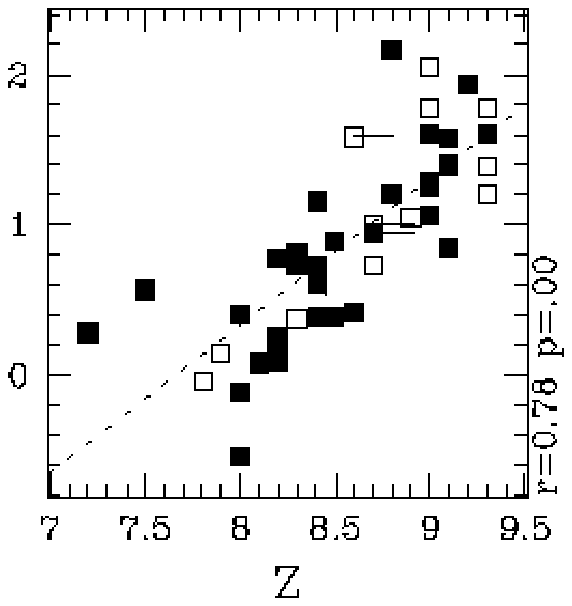}}
\vspace{15pt}
\caption{
Plots of the log of the starburst Oxygen abundance (on a scale
where the solar value is 8.9) {\it versus} two dust-indicators.
In the left plot - the spectral slope of the vacuum-UV continuum (as
parameterized by $\beta$, where F$_{\lambda} \alpha\
\lambda^{\beta}$). In the right plot, the log of the ratio of far-IR
to vacuum-UV flux.
Heavily-reddened and -extincted metal-rich starbursts lie to the upper
right of each plot.}
\end{figure}

These correlations have a straightforward interpretation: the
vacuum-UV radiation escaping from starbursts suffers an increasing
amount of reddening and extinction as the dust-to-gas ratio in the
starburst ISM increases with metallicity. This will be true
provided that neither the gas column density towards the starburst,
nor the fraction of interstellar metals locked into dust grains are
strong inverse functions of metallicity.
 
Interestingly, H98 also find that the amount of vacuum-UV extinction in
starbursts correlates strongly with the bolometric luminosity of
the starburst: only starbursts with L$_{bol}$ $<$ few $\times
10^{9}$ L$_{\odot}$ have colors expected for a lightly reddened
starburst and have vacuum-UV luminosities that rival their far-IR
luminosities. Starbursts that lie at or above the `knee' in the
local starburst luminosity function (L$_{bol}$ $>$ few $\times
10^{10}$ L$_{\odot}$ - cf. Soifer et al 1987) have red UV continua
($\beta \sim$ -1 to +0.4) and are dominated by far-IR emission
(L$_{IR}$ $\sim$ 10 to 100 L$_{UV}$). We also find that the amount
of vacuum-UV extinction in starbursts correlates well with the
absolute blue magnitude and the rotation speed of the galaxy
`hosting' the starburst: starbursts in more massive galaxies are
more dust-shrouded.\\

{\bf Third, the metallicity of the starburst also strongly affects
the UV spectrum.}
 
Apart from the effects of dust, a starburst's metallicity is the
single most important parameter in determining its vacuum-UV
properties. 
The properties of the vacuum-UV absorption-lines are strongly
dependent on metallicity. Figure 2 shows that both the
high-ionization
(e.g. CIV$\lambda$1550 and SiIV$\lambda$1400) and low-ionization
(e.g. CII$\lambda$1335, OI$\lambda$1302, and
SiII$\lambda$$\lambda$1260,1304) resonance absorption-lines are
significantly stronger in starbursts with high metallicity.
 
The metallicity-dependence of the high-ionization lines (noted
previously by Storchi-Bergmann, Calzetti, \& Kinney 1994) is not
surprising, given the likely strong contribution to these lines
from stellar winds. Theoretically, we expect that since stellar
winds are radiatively driven, the strengths of the vacuum-UV
stellar wind lines will be metallicity-dependent. This is confirmed
by available HST and HUT spectra of LMC and especially SMC stars
(Walborn et al 1995; Puls et al 1996).
 
Figure 2 also shows a metallicity-dependence for the strengths of the UV
absorption-lines that are of stellar-photospheric rather than
interstellar origin (we know they are not interstellar lines because they
correspond to transitions out of
highly excited states). Such lines are generally rather weak in starburst
spectra and/or blended with strong interstellar features. They
include CIII$\lambda$1175,
SiIII$\lambda$1417, CIII$\lambda$$\lambda$1426,1428,
SV$\lambda$1502, SiIII$\lambda$1892, and
FeIII$\lambda\lambda$1925,1960.
 
The weak but statistically-significant correlation between
metallicity and the strength of the low-ionization resonance lines
(which are primarily formed in the interstellar medium of the
starburst) is also unsurprising. Analyses of HST spectra (cf. Pettini \&
Lipman 1995;
Heckman \& Leitherer 1997; Sahu \& Blades 1997; Gonzalez-Delgado
et al 1998a) show that the strong interstellar lines are saturated
(highly optically-thick). In this case, the equivalent width of the
absorption-line (W) is only weakly dependent on the ionic column
density (N$_{ion}$): W $\propto$\ b[ln(N$_{ion}$/b)]$^{0.5}$, where
b is the normal Doppler line-broadening parameter. Over the range
that H98 sample well, the starburst metallicity increases by a
factor of almost 40 (from 0.08 to 3 solar), while the equivalent
widths of the strong interstellar lines only increase by an average
factor of about 2 to 3. This is consistent with the strong
interstellar lines being quite optically-thick.\\
 
{\bf Fourth, the properties of the strong interstellar absorption-
lines reflect the hydrodynamical consequences of the
starburst, and do not straightforwardly probe the
gravitational potential of the galaxy.}
 
As noted above, analyses of HST and HUT UV spectra of starbursts
imply that the interstellar absorption-lines lines are optically-
thick. Their strength is therefore determined to first-order by the
velocity dispersion in the starburst (see above). Thus, these lines
offer a unique probe of the kinematics of the gas in starbursts.
The enormous strengths of the starburst interstellar lines
(equivalent widths of 3 to 6 \AA\ in metal-rich starbursts) require
very large velocity dispersions in the absorbing gas (few hundred
km s$^{-1}$). Are these gas motions primarily due to gravity or to
the hydrodynamical `stirring' produced by supernovae and stellar
winds?
 
Both processes probably contribute to the observed line-broadening.
H98 find only a very weak (but still
statistically significant) correlation between the strengths
(widths) of the interstellar absorption-lines and the rotation-
speed of the host galaxy. The weakness of the correlation suggests
that gravity alone is not the whole story. 
The most direct evidence for a non-gravitational origin of the gas
motions comes from analyses of HST and HUT
spectra, which show that the interstellar lines are often
blueshifted by one-to-several-hundred km s$^{-1}$ with respect to
the systemic velocity of the galaxy (Heckman \& Leitherer 1997;
Gonzalez-Delgado et al 1998a,b; Sahu \& Blades 1997; Lequeux et al 1995
; Kunth et al 1998).
This demonstrates directly that the absorbing gas is flowing
outward from the starburst, probably helping to `feed' the superwinds
whose emission is readily observed in the optical, X-ray, and radio
regime (cf. Heckman, Lehnert, \& Armus 1993; Lehnert \& Heckman 1996a).

\section{Implications at High-Redshift}

The results summarized in section 2.2 have a variety of interesting
implications for the interpretation of the rest-frame-UV properties
of galaxies at high-redshift.
 
Powerful starbursts in the present universe emit almost all their
light in the far-infrared, not in the ultraviolet. Thus, an
ultraviolet census of the local universe would significantly
underestimate the true star-formation-rate and would systematically
under-represent the most powerful, most metal-rich starbursts
occuring in the most massive galaxies.
This {\it may} also be true at
high-redshift, where the current estimates of star-formation rely
almost exclusively on data pertaining to the rest-frame vacuum-UV.
For example, current samples might under-
represent young/forming massive elliptical galaxies.
 
Using the strong correlation between the vacuum-UV color of local
starbursts ($\beta$) and the ratio of far-IR to vacuum-UV light
emitted by local starbursts, Meurer et al (1997) estimate that an
average vacuum-UV-selected galaxy at high-redshift (e.g. Steidel
et al 1996; Lowenthal et al 1997) suffers 2 to 3 magnitudes of
extinction. 
 
As shown recently by Burigana et al (1997), the existing limits on
the far-IR/sub-mm cosmic background are consistent with the global
star-formation rates inferred by Meurer et al at z$>$2 due to dusty
starbursts unless the dust in these galaxies is quite cool
(T$_{dust}$ $<$ 20 K) compared to the dust in local starbursts
(T$_{dust}$ $\sim$ 30
to 60 K). This seems very unlikely, since the bolometric surface-brightnesses
of the high-redshift galaxies are similar to local starbursts (Meurer et al
1997), implying that the energy density of the radiation field that heats
the grains is similar in the two types of objects (cf. 
Lehnert \& Heckman (1996b).
 
In any case, it seems fair to conclude that
the history of star-formation in the universe at early times (z $>$
1) will remain uncertain until the effects of dust extinction are
better understood.
 
The strong correlation shown in Figure 3
between vacuum-UV color ($\beta$) and
metallicity in local starbursts - if applied naively to high-z
galaxies - would suggest a broad range in metallicity from
substantially subsolar to solar or higher and a median value of
perhaps 0.3 solar. This is somewhat higher than the mean
metallicity in the damped Ly$\alpha$ systems (the major repository
of HI gas at these redshifts), but this may be due to selection
effects: the UV-selected galaxies are the most actively star-
forming regions of galaxies, while the damped Ly$\alpha$ systems
tend to sample the outer, less-chemically-enriched parts of
galaxies or perhaps proto-galactic fragments (e.g. Pettini et al 1997).
 
It would also be interesting to use the correlations between absorption-
line strengths and metallicity in local starbursts to `guesstimate'
the metallicity of the high-z galaxies.
One prediction based on the local starbursts (H98) is that the
high-z galaxies should show a strong correlation between the
strength of the UV absorption-lines (stellar and interstellar) and
$\beta$ (the more metal-rich local starbursts are both redder and
stronger-lined).
 
As noted above, Meurer et al (1997; 1998) argue that the UV-selected
galaxies at high-redshift suffer substantial amounts of extinction.
If their proposed extinction-corrections are applied, the high-z
galaxies have very large bolometric luminosities ($\sim$ 10$^{11}$
to 10$^{13}$ L$_{\odot}$ for H$_{0}$ = 75 km s$^{-1}$ Mpc$^{-1}$
and q$_{0}$ = 0.1). Interestingly, the bolometric surface-
brightnesses of the extinction-corrected high-z galaxies are very
similar to the values seen in local starbursts: $\sim$ 10$^{10}$
to 10$^{11}$ L$_{\odot}$ kpc$^{-2}$. The high-redshift galaxies
appear to be `scaled-up' (larger and more luminous) versions of the
local starbursts. The physics behind this `characteristic' surface-
brightness is unclear (cf. Meurer et al 1997; Lehnert \& Heckman
1996b). However, it is intruiging that the implied average surface-
mass-density of the stars within the half-light radius ($\sim$
10$^{2}$ to 10$^{3}$ M$_{\odot}$ pc$^{-2}$) is quite similar to the
values in present-day elliptical galaxies. Are we witnessing the
formation of elliptical and/or bulges?
 
Finally, based on local starbursts - it seems likely that the gas
kinematics that are measured in the high-z galaxies using the
interstellar absorption-lines are telling us a great deal about the
hydrodynamical consequences of high-mass star-formation on the
interstellar medium, but rather little (at least directly) about
the gravitational potential or mass of the galaxy. Even the widths
of the nebular emission-lines in local starbursts are not always
reliable tracers of the galaxy potential well (cf. Lehnert \&
Heckman 1996b). This means that it will be tricky to determine
masses for the high-z galaxies without measuring real rotation-
curves via spatially-resolved spectroscopy.
 
On the brighter side, if the kinematics of the interstellar
absorption-lines can be generically shown to arise in outflowing
metal-enriched gas, we can then directly study high-redshift star-
forming galaxies caught in the act of `polluting' the intra-cluster
medium and inter-galactic medium with metals in the early universe.
 
In fact, there is now rather direct observational evidence that
this is the case. I will need to briefly digress to explain this
evidence. As emphasized above, the interstellar absorption-lines
are significantly blue-shifted with respect to the systemic
velocity of the galaxy (v$_{sys}$) in many local starbursts. In the
high-redshift galaxies there is rarely a good estimator of
v$_{sys}$ (although the weak stellar photospheric lines shown in
Figure 2 above are a promising possibility in spectra with
adequate signal-to-noise). It is also the case in local starbursts
that the true galaxy systemic velocity lies between the velocity
of the UV interstellar absorption-lines and the Ly$\alpha$ {\it
emission} line (Lequeux et al 1995; Gonzalez-Delgado et al 1998a,b;
Kunth et al 1998).
This is due to outflowing gas that both produces the blue-shifted
absorption-lines and absorbs-away the blue side of the Ly$\alpha$
emission-line. Thus, a purely-UV signature of outflowing gas is a
blueshift of the interstellar absorption-lines with respect to the
Ly$\alpha$ emission-line (even though neither is at v$_{sys}$).
 
Recently, Franx et al (1997) have seen just this effect in a spectrum
of the most distant known object in the universe: a
gravitationally-lensed galaxy at z = 4.92. They find that the
Ly$\alpha$ emission line is redshifted by about 400 km s$^{-1}$
relative to the SiII$\lambda$1260 interstellar absorption-line
across the entire face of the galaxy. More generally, Lowenthal et
al (1997) have constructed a composite UV spectrum of 12 high-z
galaxies. This also shows
a strong redshifted Ly$\alpha$ emission-
line and weak blue-shifted absorption in Ly$\alpha$ and the resonance lines
of metal ions in absorption. This
composite spectrum strongly suggests that the outflow of metal-
enriched gas at velocities of a few hundred km s$^{-1}$ is a
generic feature of the high-z galaxies. {\it If} the outflowing gas
escapes into the IGM, such flows could bring an IGM with
$\Omega_{IGM}$ $\sim$ 0.01 h$^{-2}$ up to a mean metallicity of
$>$ 10$^{-2}$ solar by a redshift of 2.5 (cf. Madau \& Shull 1996).

\section{UV Probes of the Starburst-AGN Connection in Seyfert Nuclei}

As noted in the introduction, understanding the role of starbursts
in the AGN phenomenon is one of the crucial issues in extragalactic
astronomy. I have also documented in this review the utility of UV spectroscopy
as a probe of the hot stars that power a starburst.

In order to test the `starburst-AGN connection' via UV
spectroscopy, it is clearly preferable to focus on those objects
in which the obscuring torus has providentially blocked out the
blinding glare from the central engine (thereby allowing us to
study the fainter surrounding `circumnuclear' region without
squinting). Thus, we have focused our attention on the 
nuclei of type 2 Seyfert galaxies.
 
Type 2 Seyfert nuclei have long been known to exhibit a
`featureless continuum' (`FC') that produces the UV light and
typically 10\% to 50\% of the visible/NIR light (the rest appears
to be light from an ordinary old population of bulge stars). Until
recently, it was thought that the optical/UV FC was light from the
hidden type 1 Seyfert nucleus that had been reflected into our
line-of-sight by warm electrons and/or dust. However,
recent optical spectropolarimetry (Tran 1995) shows that this is
not the case: most of the optical FC must have some other origin,
since it is significantly less polarized than the reflected broad
emission-lines. Similarly, we (Heckman et al 1995 - hereafter H95)
examined the vacuum UV (1200 to 2000~\AA) spectral properties of
a large sample of type 2 Seyferts using IUE data and showed that
the lack of any detectable reflected emission from the BLR meant
that at least 80\% of the UV continuum must have some other origin.
 
Two possibilities for the optical/UV FC have been advanced. The
first (Tran) is that the unpolarized (unreflected) component of the
FC (the `FC2') is produced by thermal emission from hot (10$^{5}$
to 10$^{6}$ K) gas heated in some way by the central energy source.
The second (Cid Fernandes \& Terlevich 1995) ascribes the FC2 to
a dust-shrouded starburst possibly associated with the obscuring
torus. The weakness of the observed line UV emission led H95 to
rule out a primarily nebular origin for the UV continuum, and to
argue instead that the `dusty starburst' interpretation is the more
plausible one (at least for the UV-bright members of the H95
sample). Moreover, we also showed that the starburst hypothesis can
be correct only if the bulk of the observed far-IR continuum from
these type 2 Seyferts is re-radiated starburst light. The large
far-IR luminosities would then imply that starbursts are an
energetically significant component of the Seyfert phenomenon.
 
However, the IUE data do not have adequate signal-to-noise
to unambiguously detect stellar absorption-lines in the Seyferts. 
Thus, they provide only
indirect and inconclusive evidence for a starburst component in the
FC of type 2 Seyferts. To directly test this requires the
spectroscopic detection of an unusually luminous population of
young stars, and this is best accomplished in the vacuum UV using
strong stellar wind lines and weaker stellar photospheric lines
(cf. Leitherer, Robert, \& Heckman 1995; Heckman \& Leitherer
1997). With data of suitable quality, a young stellar component may
also be detected in the near UV via high-order Balmer absorption-
lines and the Balmer edge from hot stars (e.g. Gonzalez-Delgado,
Leitherer, \& Heckman 1997).
%
\begin{figure}
\centerline{\epsfig{figure=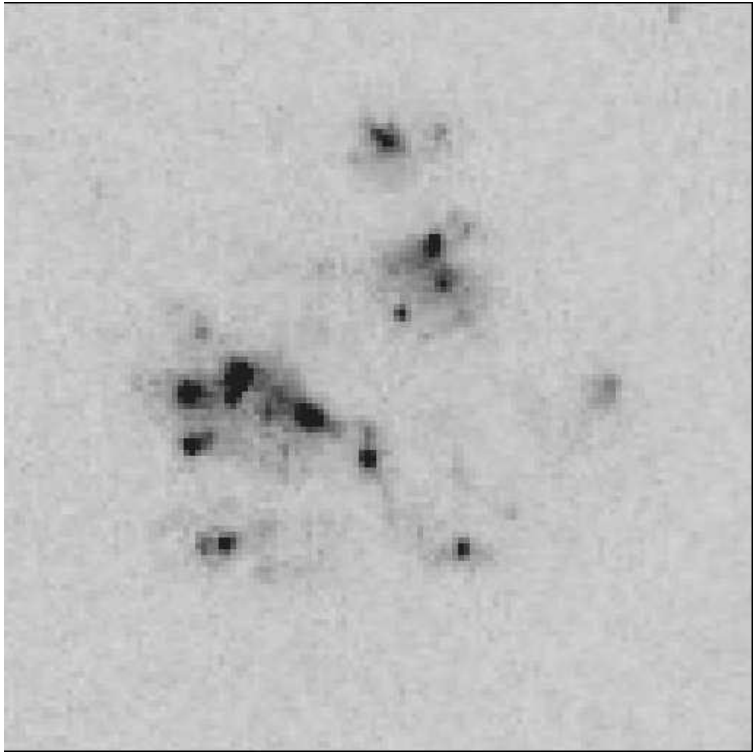}}
\vspace{15pt}
\caption{
HST FOC image of the vacuum-UV (2200~\AA) continuum emission
in the nucleus of the type 2 Seyfert galaxy NGC 5135. The region displayed
is about 1 kpc across. This image is morphologically-similar to
those of typical starbursts, including the presence of bright UV knots
corresponding to super star clusters.
(Meurer et al 1995).
}
\end{figure}
\begin{figure}
\centerline{\epsfig{figure=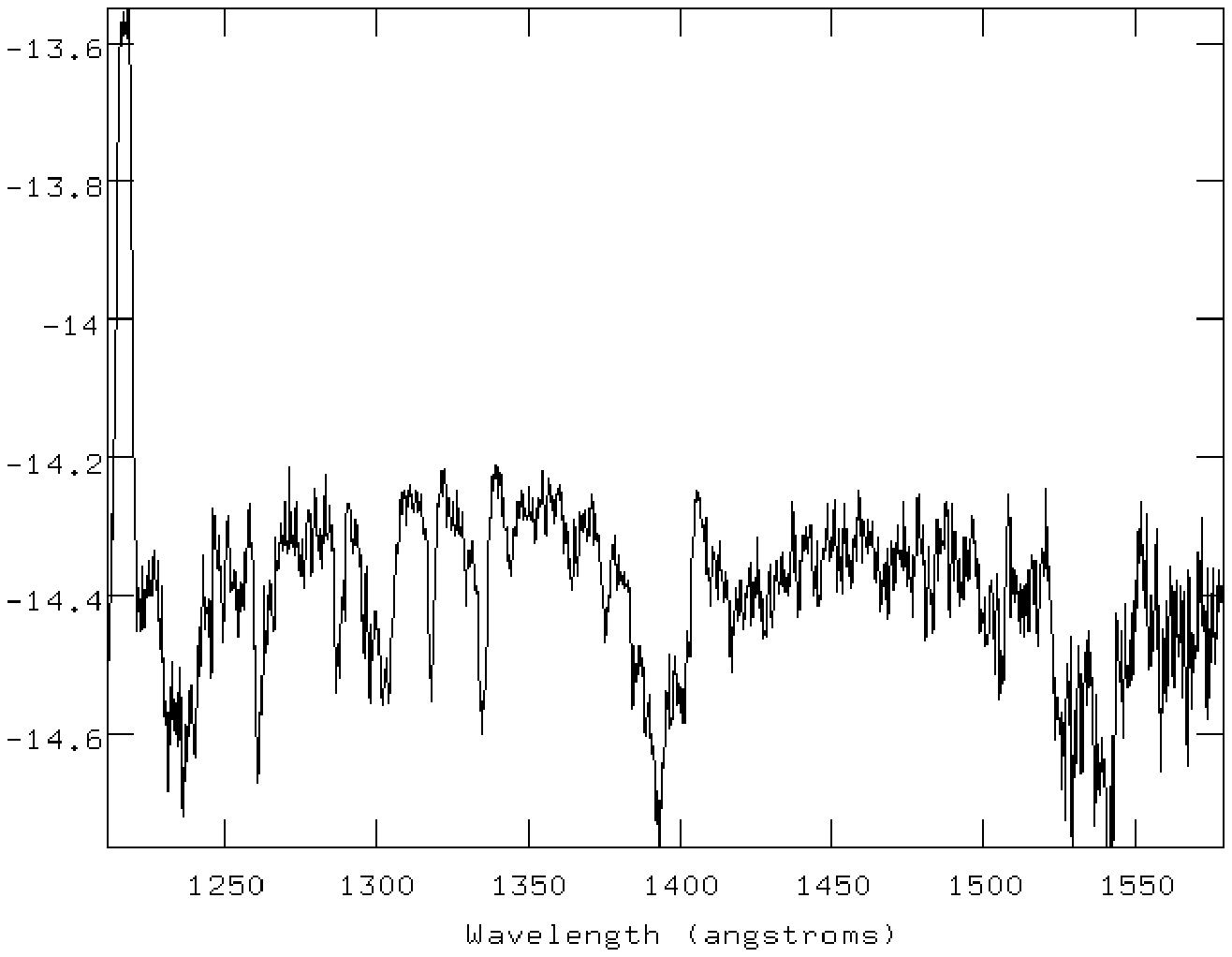}}
\vspace{15pt}
\caption{HST GHRS vacuum-UV spectrum of the
the nucleus of the type 2 Seyfert galaxy NGC 5135 (logF$_{\lambda}$ vs.
$\lambda$). This spectrum was
obtained of the region shown in Figure 4. Note the broad, blueshifted
stellar wind lines due to CIV$\lambda$1550, SiIV$\lambda$1400, and
NV$\lambda$1240 and the unshifted stellar photospheric line due to
the excited SiIII$\lambda$1299 and SV$\lambda$1502 multiplets. The other
strong narrow lines
are interstellar in origin (see Figure 2).
Comparing this spectrum to starburst spectra,
the only noticable difference is the strong Ly$\alpha$
emission-line in NGC 5135, which is most likely excited by the ionizing
radiation from the hidden `central engine'.}
\end{figure}
We have therefore undertaken a program to obtain high-resolution
vacuum UV images and spectra (with 
HST) and near UV spectra (with ground-based telescopes) of a
representative sample of the brightest type 2 Seyfert nuclei.
These results have been presented in detail in Heckman et al (1997)
and Gonzalez-Delgado et al (1998b).
HST imaging shows
that the UV continuum source in every case is spatially-resolved
(scale size few hundred pc or greater). In some cases the
morphology is strikingly reminiscent of UV images of starbursts
(compare Fig. 4 to images in Meurer et al 1995). In other cases
(cf. Capetti et al 1996), a component of the UV continuum is
roughly aligned with the inferred polar axis of the obscuring torus
(as expected for reflected and/or reprocessed light from the
central engine).
 
Of the sample of 13 type 2 Seyfert with HST vacuum-UV images, only
four are bright enough for us to obtain spectra of adequate quality
in the crucial UV spectral window from about 1200 to 1600~\AA.
However, these spectra are decisive: all four show the clear
spectroscopic signature of a starburst population that dominates
the UV continuum (Fig. 5). In addition to classic strong stellar
wind features (NV$\lambda$1240, SiIV$\lambda$1400, and
CIV$\lambda$1550), we can also detect weaker and much narrower
absorption features from  excited transitions (which are
therefore indisputably of stellar origin - cf. Heckman
\& Leitherer 1997; Heckman et al 1997; Gonzalez-Delgado et al 1998b).

In each of the four cases, if we use the empirical `starburst
attenuation law' (Calzetti et al 1994) to
correct the observed UV continuum for dust extinction, we find that
the bolometric luminosity of the nuclear (10$^{2}$-pc-scale) 
starburst is comparable to the estimated bolometric luminosity
of the `hidden' type 1 Seyfert nucleus (of-order 10$^{10}$
L$_{\odot}$). The large-aperture IUE spectra imply 
the existence of a surrounding larger-scale (few kpc)
and more powerful (few $\times$ 10$^{10}$ to 
10$^{11}$ L$_{\odot}$) dusty starburst that is energetically capable
of powering the bulk of the observed far-IR emission from the galaxy.
Thus, starbursts are an energetically
significant (or even dominant) component of at least {\it some}
Seyfert galaxies.
 
However, we have HST spectra of only four type 2 Seyferts, and
these are strongly biased in favor of cases with high UV surface-
brightness. Can we say anything more general? To address this, we
have embarked on a program to obtain spectra from about 3500 to
9000~\AA\ of a complete sample of the 25 brightest type 2 Seyfert
nuclei in the local universe. These objects are selected from
extensive lists of known Seyfert galaxies on the basis of the flux
of either the nebular line-emission from the Narrow Line Region
(the [OIII]$\lambda$5007 line) or of the nuclear radio source
(Whittle 1992).
 
We are still analysing these spectra, but even a cursory inspection
of the near UV region (below 4000~\AA) shows that at least one-third have
pronounced Balmer absorption-lines whose strength is consistent
with a population of late O or early B stars. This group includes
three of the four objects with HST vacuum UV spectra. In most of
the remainder of the sample, the `FC' is so weak relative
to the light from a normal old-bulge stellar population that its
origin is still not clear. There are also several cases in which
the Balmer emission-lines from the NLR are so strong in the near-
UV that they overwhelm any putative stellar absorption features.
Fortunately, the HST vacuum-UV spectrum of one of these (Mrk 477)
leaves no doubt that it contains a starburst (Heckman et al 1997).
 
Thus, when we complete our analysis of these data we will be in a
position to judge the general importance of circumnuclear
starbursts in the Seyfert phenomenon.
\newpage

\section{Conclusions}

In a nutshell:\\[1mm]

{\bf * Starbursts can teach us valuable lessons about the high-redshift
universe.}\\[1mm]

{\bf * Starbursts may play a key role in the AGN phenomenon.}\\[1mm]

{\bf * Good UV spectroscopic data are essential for understanding
starbursts and for probing the starburst-AGN connection. This
is true in the universe of the `here and now' and the `there and then'!}

\begin{acknow}
I would like to thank the organizers of this most enjoyable conference.
I would also like to thank my collaborators on the work described in this
paper, and especially Carmelle Robert, Claus Leitherer, Rosa Gonzalez-Delgado,
and Gerhardt Meurer. Thanks to Daniela Calzetti, Mark Dickinson,
Roberto Terlevich, and Ski Antonucci for stimulating and illuminating
discussions. This work was supported by NASA grants NAGW-3138, GO-5944,
and GO-6539.
\end{acknow}


\begin{references}

\bibitem[]{} Burigana, C., Danese, L., De Zotti, G., Franceschini,
A., Mazzei, P., \& Toffolati, L. 1997, MNRAS, 287, L17

\bibitem[]{} Calzetti, D. Kinney, A., \& Storchi-Bergmann, T. 1994,
ApJ, 429, 582

\bibitem[]{} Calzetti, D. Kinney, A., \& Storchi-Bergmann, T. 1996,
ApJ, 458, 132

\bibitem[]{} Capetti, A., Axon, D., Macchetto, F.D., Sparks, W., and Boksenberg,
A. 1996, ApJ, 466, 169

\bibitem[]{} Cid Fernandes, R., and Terlevich, R. 1995, MNRAS, 272, 423

\bibitem[]{} Conti, P., Leitherer, C., \& Vacca, W. 1996, ApJ, 461,
L87

\bibitem[]{} Fanelli, M., O'Connell, R., \& Thuan, T. 1988, ApJ, 334, 665

\bibitem[]{} Franx, M., Illingworth, G., Kelson, D., van Dokkum,
P., \& Tran, K.-V. 1997, ApJL, 486, L75

\bibitem[]{} Gallego, J., Zamorano, J., Aragon-Salamanca, A., \&
Rego, M. 1995, ApJL, 445, L1 

\bibitem[]{} Gonzalez-Delgado, R., Leitherer, C., Heckman, T., \& Cervino, M.
1997, ApJ, 483, 705

\bibitem[]{} Gonzalez-Delgado, R., Leitherer, C., Heckman, T.,
Ferguson, H., \& Lowenthal, J. 1998a, ApJ, in press

\bibitem[]{} Gonzalez-Delgado, R., Heckman, T., Leitherer, C.,
Meurer, G., Kinney, A., Koratkar, A., Krolik, J., \& Wilson, A.
1998b, submitted to ApJ

\bibitem[]{} Heckman, T. 1997, in ``Cosmic Origins of Galaxies, Planets,
and Life'', ed. J.M Shull, C. Woodward, and H. Thronson, ASP

\bibitem[]{} Heckman, T., \& Leitherer, C. 1997, AJ, 114, 69

\bibitem[]{} Heckman, T., Lehnert, M., \& Armus, L. 1993, in ``The
Evolution of Galaxies and their Environments'', Ed. M. Shull and H.
Thronson, Kluwer, 455

\bibitem[]{} Heckman, T., Krolik, J., Meurer, G., Calzetti, D., Kinney, A.,
Koratkar, A., Leitherer, C., Robert, C., and Wilson, A. 1995, ApJ,
452, 549
 
\bibitem[]{} Heckman, T., Gonzalez-Delgado, R., Leitherer, C., Meurer, G.,
Krolik, J., Wilson, A., Koratkar, A., and Kinney, A. 1997, ApJ,
482, 114

\bibitem[]{} Heckman, T., Robert, C., Leitherer, C., Garnett, D., and
van der Rydt, F. 1998, submitted to ApJ

\bibitem[]{} Huchra, J. 1977, ApJS, 35, 171

\bibitem[]{} Kinney, A., Bohlin, R., Calzetti, D., Panagia, N., \&
Wyse, R. 1993, ApJS, 86, 5

\bibitem[]{} Kormendy, J., and Richstone, D. 1995, ARA\&A, 33, 581

\bibitem[]{} Kunth, D., Mas-Hesse, J., Terlevich, E., Terlevich,
R., Lequeux, J., and Fall, S.M. 1998, A\&A, in press

\bibitem[]{} Lehnert, M., \& Heckman, T. 1996a, ApJ, 462, 651

\bibitem[]{} Lehnert, M., \& Heckman, T. 1996b, ApJ, 472, 546

\bibitem[]{} Leitherer, C. 1997, in ``The Ultraviolet Universe at Low
and High Redshift: Probing the Progress of Galaxy Evolution'', ed. W. Waller,
M. Fanelli, J. Hollis, and A. Danks (AIP: Woodbury, NY), p. 119

\bibitem[]{} Leitherer, C., Walborn, N., Heckman, T., \& Norman, C.
1991, ``Massive Stars in Starburst Galaxies'', Cambridge University Press

\bibitem[]{} Leitherer, C., Robert, C., \& Heckman, T. 1995, ApJS,
99, 173

\bibitem[]{} Leitherer, C., \& Heckman, T. 1995, ApJS, 96, 9

\bibitem[]{} Leitherer, C., Vacca, W., Conti, P., Filippenko, A.,
Robert, C., \& Sargent, W. 1996, ApJ, 465, 717

\bibitem[]{} Lequeux, J., Kunth, D., Mas-Hesse, J.,\& Sargent, W.
1995, A\&A, 301, 18

\bibitem[]{} Lowenthal, J., Koo, D., Guzman, R., Gallego, J.,
Phillips, A., Faber, S., Vogt, N., \& Illingworth, G. 1997, ApJ, 481, 673

\bibitem[]{} Madau, P., \& Shull, S.M. 1996, ApJ, 457, 551

\bibitem[]{} Madau, P., Ferguson, H., Dickinson, M., Giavalisco,
M., Steidel, C., \& Fruchter, A. 1996, MNRAS, 283, 1388

\bibitem[]{} Meurer, G., Heckman, T., Leitherer, C., Kinney, A.,
Robert, C., \& Garnett, D. 1995, AJ, 110, 2665

\bibitem[]{} Meurer, G., Heckman, T., Leitherer, C., Lowenthal, J.,
\& Lehnert, M. 1997, AJ, 114, 54

\bibitem[]{} Meurer, G., Heckman, T., \& Calzetti, D. 1998, in
preparation

\bibitem[]{} Miyoshi, M., Moran, J., Herrnstein, J., Greenhill, L., Nakai, N.,
Diamond, P., and Inoue, M. 1995, Nature, 373, 127

\bibitem[]{} Norman, C., and Scoville, N. 1988, ApJ, 332, 134

\bibitem[]{} Perry, J., and Dyson, J. 1985, MNRAS, 213, 665

\bibitem[]{} Pettini, M., \& Lipman, K. 1995, A\&A, 297, 63

\bibitem[]{} Pettini, M., Smith, L., King, D., \& Hunstead, R. 1997, ApJ, 486,
665

\bibitem[]{} Rees, M. 1984, ARA\&A, 22, 471

\bibitem[]{} Sahu, M., \& Blades, J.C. 1997, ApJ, 484, L125

\bibitem[]{} Sanders, D., and Mirabel, I.F. 1996, ARA\&A, 34, 749

\bibitem[]{} Sekiguchi,K., and Anderson, K. 1987, AJ, 94, 644

\bibitem[]{} Soifer, B.T., Sanders, D., Madore, B., Neugebauer, G.,
Lonsdale, C., Persson, S.E., \& Rice, W. 1987, ApJ, 320, 238

\bibitem[]{} Steidel, C., Giavalisco, M., Pettini, M., Dickinson, M.,
\& Adelberger, K. 1996, ApJ, 462, L17

\bibitem[]{} Storchi-Bergmann, T., Kinney, A.L. \& Challis, P. 1995, ApJS, 98,
 103

\bibitem[]{} Tanaka, Y. Nandra, K., Fabian, A., Inoue, H., Otani, C. Dotani, T.,
Hayashida, K., Iwasawa, K., Kii, T., Kunieda, H., Makino, F., and
Matsuoka, M., 1995, Nature, 375, 659

\bibitem[]{} Terlevich, R., and Melnick, J. 1985, MNRAS, 213, 841

\bibitem[]{} Tran, H. 1995, ApJ, 440, 597

\bibitem[]{} Weedman, D. 1983, ApJ, 266, 479

\bibitem[]{} Whittle, M. 1992, ApJS, 79, 49

\end{references}
\end{document}